\documentclass{iaus}
\usepackage{graphicx}

\title[Various applications of  multicolour photometry and radial velocity data] 
{Various applications of  multicolour photometry and radial
       velocity data\\ for multimode $\delta$ Scuti stars}

\author[J. Daszy\'nska-Daszkiewicz, W.A. Dziembowski, A.A. Pamyatnykh et al.]   
{J. Daszy\'nska-Daszkiewicz$^{1,2}$, W. A. Dziembowski$^{2,3}$, \\A. A. Pamyatnykh$^{2,4}$
M. Breger$^5$, W. Zima$^5$}

\affiliation{$^1$Astronomical Institute of Wroc{\l}aw University,
ul. Kopernika 11, Poland, \break email: daszynska@astro.uni.wroc.pl\\[\affilskip]
$^2$Copernicus Astronomical Center, ul. Bartycka 18, 00-716 Warsaw, Poland\\[\affilskip]
$^3$ Warsaw University Observatory, Al. Ujazdowskie 4, Poland\\[\affilskip]
$^4$ Institute of Astronomy, Russian Academy of Science, Pyatnitskaya 48, Moscow, Russia\\[\affilskip]
$^5$ Institute of Astronomy, University of Vienna, T\"urkenschanzstr. 17, Austria}

\pubyear{2004}
\volume{224}  
\pagerange{119--126}
\date{?? and in revised form ??}
\jname{The A-Star Puzzle}
\editors{J.\,Zverko, W.W.\,Weiss, J.\,\v{Z}i\v{z}\v{n}ovsk\'{y}, \& S.J.\,Adelman, eds.}
\begin{document}

\maketitle

\begin{abstract}
In addition to revealing spherical harmonic degrees, $\ell$, of
excited modes, pulsational amplitudes and phases from multicolour
photometry and radial velocity data yield a valuable  constraints
on stellar atmospheric parameters and on subphotospheric
convection. Multiperiodic pulsators are of particular interest
because each mode yields independent constraints.  We present an
analysis of data on twelve modes observed in FG Vir star.
\end{abstract}

\firstsection 

\section{Introduction}

In a recent paper Daszy\'nska-Daszkiewicz, Dziembowski, Pamyatnykh (2003, Paper I)
showed that photometric amplitudes and phases of pulsating stars
are useful not only to identify the spherical harmonic degree,
$\ell$, but also for constraining models on stellar convection and
stellar atmospheric parameters.

To calculate theoretical values of photometric amplitudes and
phases we have to make use of the complex nonadiabatic parameter,
$f$, which gives the ratio of the local flux perturbation to the
radial displacement at the photosphere. The $f$  parameter is
obtained from linear nonadiabatic calculation of oscillation in
relevant stellar models. In the case of $\delta$ Scuti star
models, $f$  is very sensitive to the treatment of subphotospheric
convection. We also need models of stellar atmospheres for
evaluation of the flux derivatives with respect to effective
temperature and gravity and the limb-darkening coefficients.
The most popular are Kurucz models (1998) but now there are
alternative models available, e.g. Phoenix (Hauschildt et al.
1997) or NEMO.2003 (Nendwich et al. 2004).

In Paper I we proposed the method of extracting simultaneously
$\ell$ and $f$ from multicolour data and applied it to several
$\delta$ Scuti stars. In all cases the $\ell$ identification were
unique and the inferred values of $f$ were sufficiently accurate
to yield a useful constraints on convection. In the follow-up
paper (Daszy\'nska-Daszkiewicz et al. 2004), we included radial
velocity measurements, which improved significantly the
determination of $\ell$ and $f$. The very important feature of
this method is that we can determine the $\ell$ values in 
$\delta$ Scuti stars avoiding major theoretical uncertainty
concerning subphotospheric convection.

Multimode pulsators are of special interest because each mode gives
us independent constraints on stellar parameters.  FG Vir is
the most multimodal $\delta$ Scuti pulsator. After last
photometric and spectroscopic campaigns the total number of
detected modes increased up to 48 (Breger et al. 2004, Zima et al.
2004). For twelve frequencies we have both two-colour Str\"omgren
photometry $(v,y)$, as well as radial velocity data. We apply our
method to these modes using various models of stellar atmospheres.

\section{Observations}

Two recent photometric campaigns on FG Vir took place in 2002 and
2003. Spectroscopic observations were carried out in 2002. In
Fig.1 we show how amplitudes and phases change from season 2002 to
season 2003.  We can see that for some of the modes, the
differences are significant. This problem is more evident if we
combine photometry and spectroscopy. In Fig.2 we present
diagnostic diagrams for the spherical harmonic degree, $\ell$,
constructed from photometric and spectroscopic observables.\
The conclusion is that only contemporaneous observations can be used.

\begin{figure}
 \includegraphics[width=\textwidth,height=9cm]{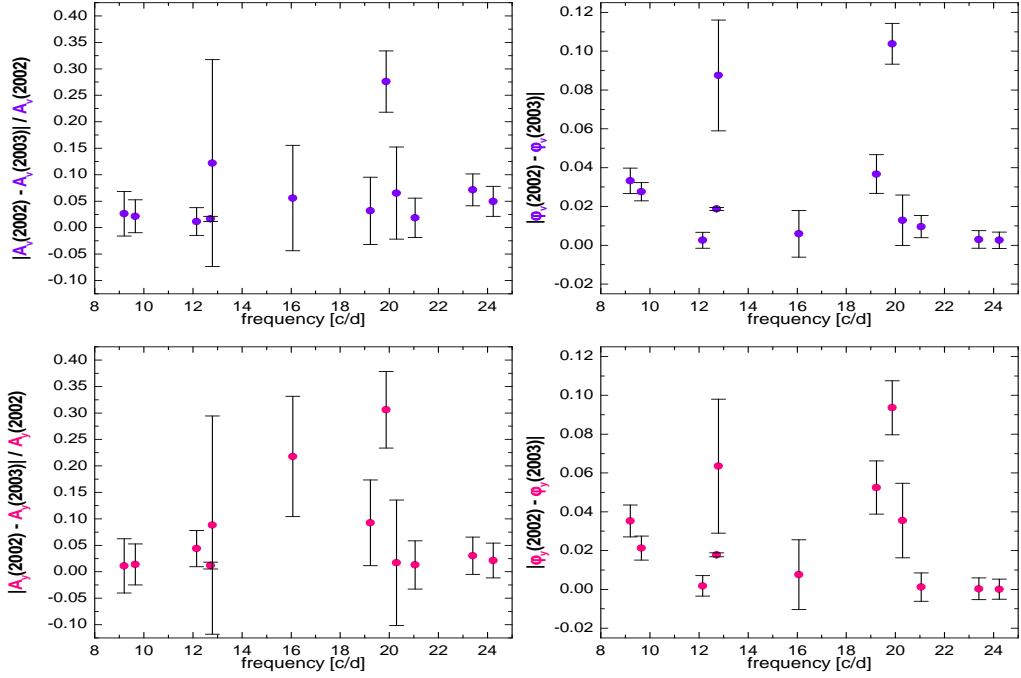}
  \caption{Changes in photometric amplitudes and phases between 2002 and 2003.}
  \label{fig:diff}
\end{figure}

\begin{figure}
 \includegraphics[width=\textwidth]{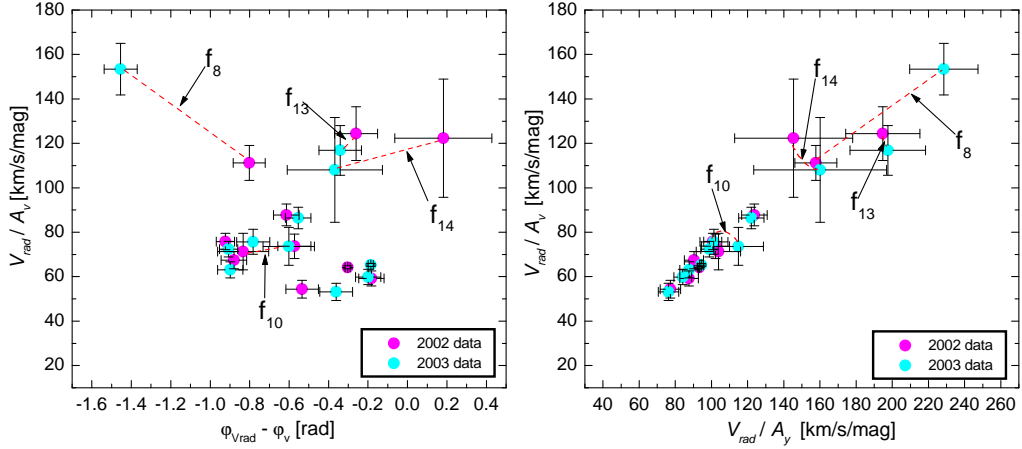}
  \caption{Diagnostic diagrams for the spherical harmonic degree, $\ell$,
          using combined photometric and spectroscopic data.}
  \label{fig:diag}
\end{figure}

\section{Method of inferring $\ell$ and $f$ from observations}

The method is described in detail in Paper I. Here we give only a brief outline.

The complex photometric amplitudes for a number of passbands, $\lambda$,
are written in the form of the linear observational equations
\begin{equation}
{\cal D}_{\ell}^{\lambda} ({\tilde\varepsilon} f)
+{\cal E}_{\ell}^{\lambda} {\tilde\varepsilon} = A^{\lambda},
\end{equation}
where
\begin{equation}
{\tilde\varepsilon} = \varepsilon Y^m_{\ell}(i,0),
\end{equation}
\begin{equation}
{\cal D}_{\ell}^{\lambda} = \frac14 b_{\ell}^{\lambda}
\frac{\partial \log ( {\cal F}_\lambda |b_{\ell}^{\lambda}| ) }
{\partial\log T_{\rm{eff}}}
\end{equation}
\begin{equation}
{\cal E}_{\ell}^{\lambda} =  b_{\ell}^{\lambda}
\left[ (2+\ell )(1-\ell ) - \left( \frac{\omega^2 R^3}{G M} + 2 \right)
\frac{\partial \log ( {\cal F}_\lambda
|b_{\ell}^{\lambda}| ) }{\partial\log g} \right]
\end{equation}
Having spectroscopic observations we can supplement the above set
with an expression for the radial velocity (the first moment, ${\cal M}_1^{\lambda}$)
\begin{equation}
{\rm i}\omega R \left( u_{\ell}^{\lambda}
+ \frac{GM}{R^3\omega^2} v_{\ell}^{\lambda} \right)
{\tilde\varepsilon}={\cal M}_1^{\lambda}
\end{equation}
Symbols in Eqs. (3.3-5) have the following meaning. $\varepsilon$
is a complex parameter fixing mode amplitude and phase, $i$ is
inclination angle, and
$b_{\ell}^{\lambda},~u_{\ell}^{\lambda},~v_{\ell}^{\lambda}$ are
limb-darkening-weighted  disc averaging factors. ${\cal F}(T_{\rm
eff},g)$ is the monochromatic flux from static atmosphere models.
Mean atmospheric parameters, ($T_{\rm eff}, \log g,$ [m/H]), enter
through ${\cal D}_{\ell}^{\lambda}$, ${\cal E}_{\ell}^{\lambda}$
and through the disc averaging factors, which contain the
limb-darkening coefficients.

In the calculations reported in this paper, we used ${\cal
F}_\lambda(T_{\rm eff}, \log g, [m/H])$ from the Kurucz, Phoenix,
and NEMO.2003 models. The limb-darkening coefficients were taken
from Claret (2000, 2003) for the Kurucz and Phoenix models, and
from Barban et al. (2003) for the NEMO.2003 models.

Each passband, $\lambda$, yields r.h.s. of equations (3.1).
Measurements of radial velocity yield r.h.s. of equation (3.5).
With data from two passbands we have three complex linear
equations for two complex unknowns: ${\tilde\varepsilon}$ and
$({\tilde\varepsilon}f)$ (note that $|\varepsilon f|$ is the
relative amplitude of the luminosity variations). The equations
are solved by LS method for specified $\ell$ values. The $\ell$
determination is based on the behavior of minima of
$\chi^2(\ell)$. The inferred $f$ values are of interest as
constraints on stellar convection models (Paper I).

\section{Application to FG Vir}

FG Vir is a well known $\delta$ Scuti variable located in the middle
of its Main Sequence evolution. The basic stellar parameters, as derived
from mean photometric indices and Hipparcos parallax, are:
$\log T_{\rm eff}=3.875\pm 0.009$, $\log g=4.0\pm 0.1$.
The mass estimated from evolutionary tracks is $M=1.85\pm 0.1 M_{\odot}$.
As was shown by Mittermayer \& Weiss (2003), FG Vir has
the solar chemical composition, thus in our calculations
we adopt the standard metal abundance $Z=0.02$.

\subsection{Identification of spherical harmonic degree, $\ell$}

We applied the method to twelve frequencies of FG Vir. All of them
have both photometric and spectroscopic data. We use only
observations from 2002 because radial velocity measurements are
only from 2002. In the present application the radial velocity data
are essential because we have data for only two photometric
passbands and three is the minimum if we want to rely on the pure
photometric version of our method.
\begin{figure}
 \includegraphics[width=\textwidth,height=11cm]{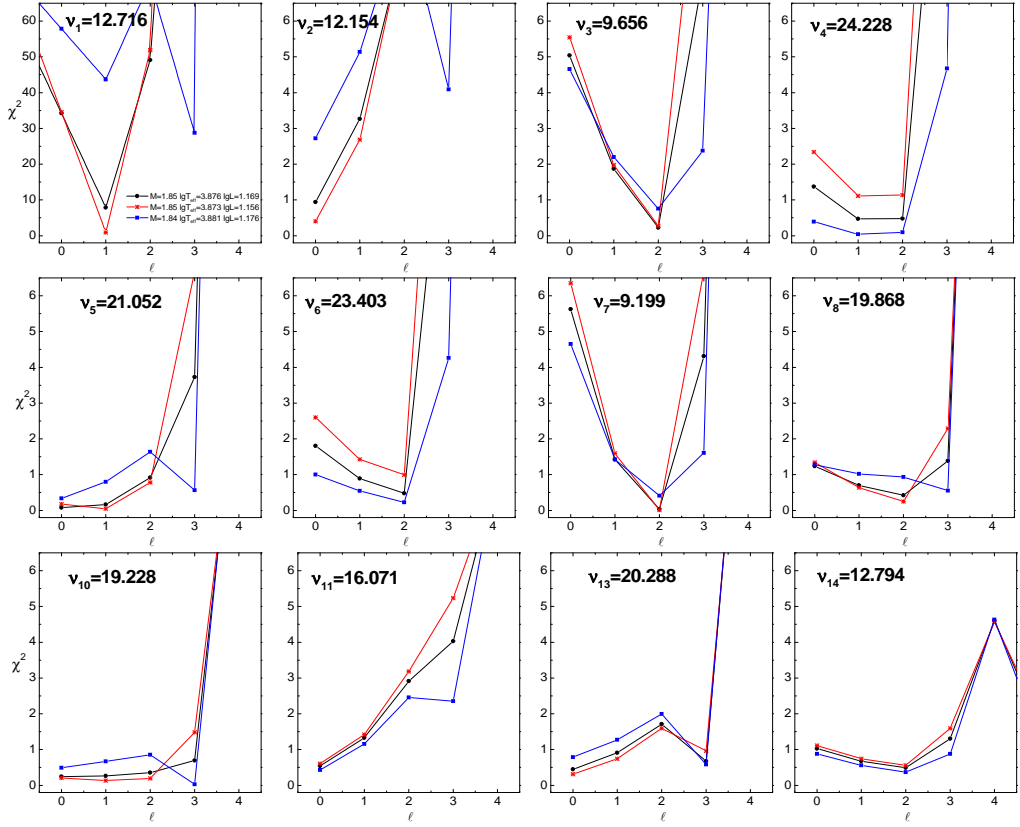}
  \caption{The values of $\chi^2$ from the method described in the text as a function of $\ell$.}
 \label{fig:lident}
\end{figure}
\begin{table}\def~{\hphantom{0}}
  \begin{center}
  \caption{Possible identification of $\ell$ ($\chi^2\le 1.6$) within the accepted $T_{\rm eff}$ range}
  \label{tab_ident}
  \begin{tabular}{lcccc}\hline
   $\nu$ [c/d] & our identification & Breger et al. & Viscum et al. & Breger et al.\\
               & from phot.\&spec.  &     (1995)    &    (1998)     &     (1999)   \\
      \hline
   $\nu_1$= 12.7164   & $\ell=1$  & $\ell=0$ & $\ell=1$& $\ell=1$ \\
   $\nu_2$= 12.1541   & $\ell=0$  & $\ell=2$ & $\ell=0$& $\ell=0$ \\
   $\nu_3$=  9.6563   & $\ell=2$  & $\ell=2$ & $\ell=2$& $\ell=1,2$ \\
   $\nu_4$= 24.2280   & $\ell=2,1,0$  & $\ell=0$ & $\ell=1$& $\ell=1,2$ \\
   $\nu_5$= 21.0515   & $\ell=1,0,2$  & $\ell=1$ & $\ell=2$& $\ell=2$ \\
   $\nu_6$= 23.4033   & $\ell=2,1,0$  & $\ell=2$ & $\ell=0$& $\ell=0,1$ \\
   $\nu_7$=  9.1991   & $\ell=2,1$  & $\ell=2$ & $\ell=2$& $\ell=2$ \\
   $\nu_8$= 19.8676   & $\ell=2,1,0$  & $\ell=2$ & $\ell=2$& $\ell=2$ \\
\hline
  \end{tabular}
 \end{center}
\end{table}

In Fig. 3 we plot $\chi^2$ as a function of $\ell$. The results
were obtained using Kurucz models. We see that the discrimination
of the $\ell$'s is sometimes better, sometimes worse. A unique
$\ell$ identification is possible only for the highest amplitude
modes. In Table 1 we compare our $\ell$ identifications with earlier
ones. Clearly, we are less optimistic than our predecessors. We
rejected only the $\ell$ values leading to $\chi^2\ge 1.6$. With
this criterion we could assign a unique $\ell$ values only to
three highest peeks. It is significant that $\ell>3$ is
excluded in all twelve cases at the safe confidence level.

\subsection{Constraints on stellar convection}

In Fig. 4 we compare empirical values of parameters $f$ deduced
from the data with the values obtained from linear nonadiabatic
calculation. Here we rely on a rather naive treatment
of convection: mixing-length theory (MLT) and convective
flux freezing approximation.

The empirical $f$'s depend only weakly on adopted values of
$T_{\rm eff}$ and $L$. The plotted values were obtained for
$M=1.85 M_{\odot}$, $\log T_{\rm eff}=3.873$ and $\log L=1.156$.
These parameters are consistent with mean photometric data for FG Vir
and evolutionary models, and they lead for the three highest amplitude
modes to the lowest $\chi^2$ obtain with our method. 
The empirical $f$'s are relatively insensitive also to the
choice of $\ell$ which in some cases is ambiguous. For the plot we
use $\ell$'s corresponding to $\chi^2$ minimum. The calculated
$f$'s are even less $\ell$-dependent.

We can see not a bad agreement with theoretical values 
of the $f$ parameter for models calculated with $\alpha=0.0$. 
This is good because our approximations are relevant
in the limit of totally inefficient convection.
\begin{figure}
 \includegraphics[width=\textwidth]{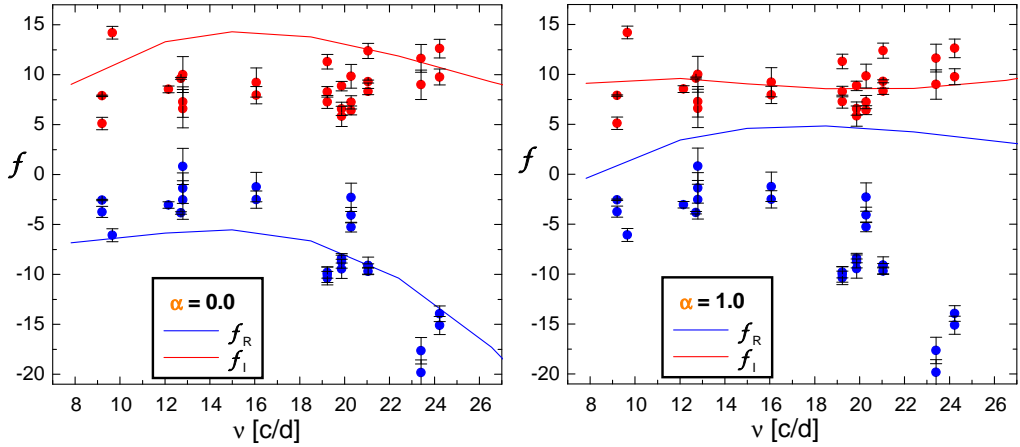}
  \caption{Comparison of empirical values of nonadiabatic parameter, $f$, (dots with error bars)
         with the theoretical ones calculated for two values of MLT parameter, $\alpha$.}
  \label{fig:fparam}
\end{figure}

\subsection{Consequences of using different models of stellar atmospheres}

Our determination of $\ell$ and $f$ requires atmospheric models.
All our results so far were obtained with the use of Kurucz
tabular data. In our comparison we use the data from Phoenix and
NEMO.2003 models. Such a comparison is important for assessing
reliability of the quoted $\ell$ and $f$ values.
 Models of stellar atmospheres are needed for evaluation coefficients
${\cal D}_{\ell}^{\lambda}$,
 ${\cal E}_{\ell}^{\lambda}$ and $b_{\ell}^{\lambda}$.
The most important quantity derived from the models is the
temperature derivative of the monochromatic flux
\begin{equation}
\alpha_T= \frac{\partial \log {\cal F}_\lambda } {\partial\log
T_{\rm eff}},
\end{equation}
which enters ${\cal D}_{\ell}^{\lambda}$. This quantity depends
most strongly on  $T_{\rm eff}$ and the same must be true for the
minimum $\chi^2$. This suggest that the best value of $T_{\rm
eff}$ in the same as the best value of $\ell$.

We found that relying on different atmospheric models, the
$\ell$ identification based on $\chi^2$ minima is unchanged but
the minimum values of $\chi^2$ in the accepted $T_{\rm eff}$
range were much larger than the ones obtained with the use of
Kurucz data. We then relaxed the temperature constraint and
considered models in a wide range of $T_{\rm eff}$. We use
the fit of $\nu_2$ to the radial fundamental mode and evolutionary
tracks to derive corresponding value of $\log g$. We encountered
two problems, which are illustrated in Fig.5:
\begin{itemize}
\item acceptable minima of $\chi^2$ for Phoenix and NEMO.2003 models
         were found outside the acceptable $T_{\rm eff}$ range
\item in Kurucz and Phoenix models the temperature flux derivatives, $\alpha_T$,
  are non-smooth which gives bad shape of $\chi^2(T_{\rm eff})$
\end{itemize}

\begin{figure}
 \includegraphics[width=\textwidth]{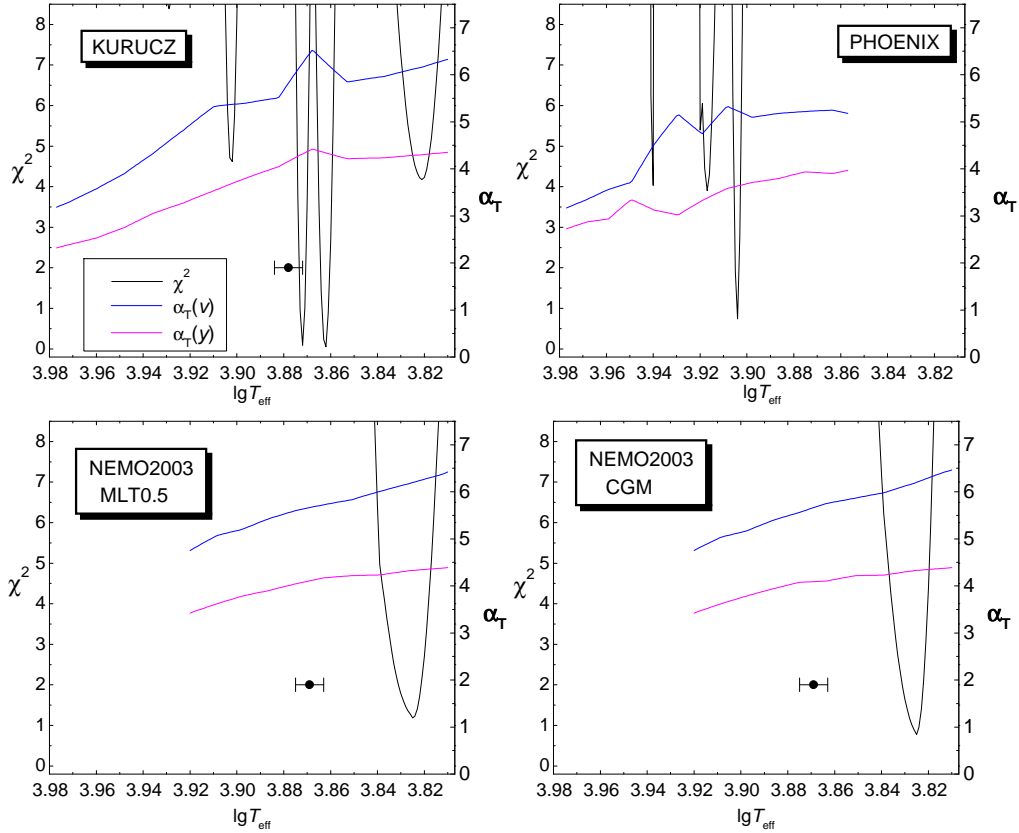}
  \caption{$\chi^2$ as a function of effective temperature a the dominant mode, $\nu_1$.
   Dots with error bars show the values derived from mean colors.}
  \label{fig:teff}
\end{figure}

\section{Conclusions}

A unique identification of spherical harmonic degree, $\ell$, of
excited modes without a priori knowledge of complex parameter $f$,
which links the surface flux variation to displacement, is
possible. However, in the case of low amplitude modes, more accurate
observations and measurements in more passbands are needed.

The values of $f$ inferred from data for FG Vir are crudely consistent
with models calculated assuming inefficient convection
$(\alpha\approx0.0)$. We showed that there is a prospect for constraining
$T_{\rm eff}$ from the pulsation data. However, for this application we
need more accurate derivatives of the fluxes in various passbands with respect to
$T_{\rm eff}$ than the derivatives we may calculated from available
tables of atmospheric models.

There are indeed various applications of amplitude and phase data
from photometric and spectroscopic observations of pulsating
stars. Multimode pulsators are advantageous, but the mode
amplitudes have to be determined with a high accuracy.
Data from photometric observations in more than two passbands
are very much desired.
\
\begin{acknowledgments}
The work was supported by KBN grant  No. 5 P03D 012 20.
\end{acknowledgments}

\end{document}